\documentclass[referee]{aa} 
\usepackage{txfonts}
\usepackage{graphicx}

\begin{document}
\title{Radio emission from the Cygnus Loop and its spectral characteristics
\thanks{Based on observations with the Effelsberg 100-m radio
telescope  operated by the Max-Planck-Institut
f\"ur Radioastronomie (MPIfR), near Bonn, Germany.}
}

\author{B. Uyan{\i}ker
          \inst{1,2},
          W. Reich\inst{1},
          A. Yar \inst{1,2},
          \and
          E. F\"urst\inst{1}}

\offprints{B. Uyan{\i}ker}

\institute{  Max-Planck-Institut f\"ur Radioastronomie, Auf dem H\"ugel 69,
             53121 Bonn, Germany\\
             email: uyaniker@mpifr-bonn.mpg.de,
             wreich@mpifr-bonn.mpg.de,
             aylin@mpifr-bonn.mpg.de,
             efuerst@mpifr-bonn.mpg.de
\and
             National Research Council, Herzberg Institute of Astrophysics,
             Dominion Radio Astrophysical Observatory,
             P.O. Box 248, Penticton, British Columbia, V2A 6K3, Canada}

\date{Received 3 December 2002 / Accepted 16 June 2004 }

\titlerunning{Radio emission from the Cygnus Loop}

\abstract {We present a new sensitive 2675~MHz radio continuum map of
the Cygnus Loop, which is used in conjunction with 408~MHz, 863~MHz
and 1420~MHz maps from both the Effelsberg 100-m telescope and the
DRAO Synthesis Telescope for a spectral analysis.  Between 408~MHz and
2675~MHz we find an overall integrated spectral index of $\rm \alpha =
-0.42 \pm 0.06$ ($S\! \sim\! \nu^{\,\alpha}$), close to previous
results.  There is no indication of a spectral break in the
integrated spectrum.  Spatially highly varying and rather strong
spectral curvature was previously reported, but is not confirmed on
the basis of new, higher sensitivity observations.  We found spectral
variations across the Cygnus Loop reaching up to $\Delta\alpha = 0.2$
from a TT-plot analysis.  The flattest spectra are seen towards
enhanced emission areas.  Spectral index maps produced between
different frequency pairs, as well as all four maps, revealed that
there are at least three flat spectrum regions.  In regions interior
to the high emission filaments, we have detected at least two spectral
components across the whole object with $\alpha=-0.4$ and
$\alpha=-0.6$ towards northern and southern parts of the object,
respectively.
 }
\maketitle

%

\section{Introduction}

The Cygnus Loop is an intense nearby non-thermal radio source, which
has been frequently studied the early years of radio astronomy
until now. Due to its strength, large size and its location at high
Galactic latitude with little confusion by unrelated emission along
the line of sight, observations were made across the entire accessible
electro-magnetic spectrum. Although the Cygnus Loop is often
considered as a typical shell-type supernova remnant (SNR) of medium
age, its morphology deviates largely from a spherical shell.  From a
recent analysis of the Cygnus Loop region using sensitive 2675~MHz
continuum and polarization data in conjunction with optical, X-ray and
infrared data, Uyan{\i}ker et al.~(\cite{uya02}) concluded that the
Cygnus Loop consists of two supernova remnants rather than
being a single SNR shaped by strong inhomogeneities in the
interstellar medium.  The significantly polarized ``break-out'' region
towards the south originates from one supernova remnant, which is
separate but interacting with the X-ray bright big northern shell.
The anomalous X-ray source AX J2049.6+2939 detected by the ASCA
satellite (Miyata et al.~\cite{miy01}) is located at the center of the
southern remnant, probably the neutron star left from the supernova
explosion.

The large and bright Cygnus Loop, comprising the two remnants
G74.3$-$8.4 and G72.9$-$9.0, is an ideal object for detailed spectral
index studies, which may reveal interesting details of the internal
SNR physics and mutual influence with the interstellar medium, as well
as possible differences between the two SNRs and their interaction.
The spectral index $\rm\alpha$, related to the flux density $S_\nu$
and the frequency $\nu$ for power law spectra
($S_{\nu}\!\!\sim\!\nu^{\,\alpha}$), reflects the conditions in the
compression regions formed by the interaction of the shock wave of the
SNR with the interstellar medium of the environment. The majority of
shell-type SNRs seem to be in the adiabatic phase. Particles
accelerated in the strong compression of the shock wave result in a
radio spectrum of $\alpha \simeq -0.5$.  A steep spectrum of less
than about $\alpha \simeq -0.7$ is a sign for a young SNR, whose
emission is dominated by the SN ejecta. Old remnants also have steep
spectra, but for another reason: particle acceleration at later stages
of evolution becomes inefficient and the spectrum reflects that of the
Galactic synchrotron background.

\begin{figure*}\centering 
\vspace{8cm}
\caption{Total intensity map of the Cygnus Loop at 2675~MHz as
observed with the Effelsberg 100m telescope. The HPBW is 4\farcm 3 .
The lowest contour on the left panel is at 10 mK T$_{\rm B}$.
Next contours start at 50 mK T$_{\rm B}$ and run in
100 mK T$_{\rm B}$ steps.
}
\label{i2675}
\end{figure*}

A spatially resolved spectral analysis between 408~MHz and 2695~MHz
was made by Green~(\cite{gre90}) revealing some small but systematic
variations between the northern and southern part of the Cygnus Loop.
A more recent three frequency study (408~MHz, 1420~MHz and 2695~MHz)
by Leahy \& Roger~(\cite{lea98}) results in larger local variations of
the spectral index including frequency dependent positive and negative
curvature in some areas of the Cygnus Loop.  This finding, if
confirmed, makes the Cygnus Loop even more interesting compared to the
majority of shell-type SNRs, where such a phenomenon is rare and
generally very weak.  Varying spectral indices on small scales across
a SNR have far reaching consequences for their internal physics, their
evolution and also on their interaction with the interstellar medium.
But the rarity of this phenomenon might also be the result of
insufficient spatial resolution that smooths out such details.

The large size of the Cygnus Loop, about $3\fdg 5 \times 2\fdg 5$,
while facilitating elaborate spectral analysis, however, require to
take care of a number of instrumental limitations. For instance, the
determination of background level due to the extent of the emission
becomes difficult and mosaicking of several images is often needed. In
addition, diffuse Galactic emission, which has a steeper spectrum
(above $\sim400$~MHz) than most shell-type SNRs, may vary across the
field and its gradient might influence the SNR spectrum in particular
for weak diffuse emission regions.  This is also noticeable in
difficulties to determine an accurate overall integrated spectrum,
where inconsistent scaling factors and low resolution of some earlier
maps add further difficulties on top of some inconsistent baselevels
between some measurements. Thus the previously suggested break at
about 1~GHz in the integrated spectrum of the Cygnus Loop (e.g. Kundu
\& Becker~\cite{kun72}; DeNoyer~\cite{den74}) must be reconsidered on
the basis of new higher quality measurements.

Significant spectral variations are not observed in most SNRs. These
are generally unexpected when a straight power law fits the spectrum
of the integrated emission quite well. Just a few SNRs show some
bending in the sense of steepening of the integrated spectrum towards
higher frequencies.  The physical reason for this bending is not quite
clear and for most objects it is observationally challenging to
measure the extent of these local spectral variations, since not all
SNRs are resolved sufficiently well by single-dish telescopes.  On the
other hand, synthesis telescopes have sufficient angular resolution
but might have problems to ``see'' the same spatially filtered
structures at different frequencies and often miss some large-scale
emission if not properly corrected by single-dish data.

\begin{figure*} 
\vspace{8cm}
\caption{1400~MHz map at $9\farcm 35$ angular resolution made with the
Effelsberg 100-m telescope (Uyan{\i}ker et al.~\cite{uya99}),
reprocessed without Stockert data.  The lowest contours on the left
panel are at 100 and 200 mK T$_{\rm B}$. The following contours start
at 500 mK T$_{\rm B}$ and run in 250 mK T$_{\rm B}$ steps.}
\label{i1420}
\end{figure*}

In this paper we analyze the spectral characteristics of the Cygnus
Loop, on the basis of higher quality data, in an attempt to clarify
the case of the reported spectral break in the integrated spectrum and
especially the unusual spatial spectral index variations across the
whole source showing both negative and positive spectral curvature as
a function of frequency.  The paper is organized as follows: in
Section 2 we describe the Effelsberg 2675~MHz observations and the
reprocessing of the archival DRAO 408~MHz and 1420~MHz data. Radio
appearance of the Cygnus Loop is discussed in Section 3 and a spectral
analysis of the Cygnus Loop using different methods and including a
recent 863~MHz Effelsberg map is presented in Section 4.  In Section 5
we analyze and discuss the results of the spectral analysis along with
a brief theoretical background on that issue, and compare our results
with previous investigations.

\section{Observations}

\begin{figure*} 
\vspace{8cm}
\caption{Reprocessed 408~MHz maps of the Cygnus Loop from the DRAO
archive.  Short spacings are provided using the Haslam et
al.~(\cite{has82}) all-sky survey data. The HPBW is $3\farcm
4\times6\farcm 9$ (EW $\times$ NS). The contours on the left
panel start at 40 K T$_{\rm B}$ and run in 8 K T$_{\rm B}$ steps.
A gradient of background emission contributed from the low resolution
data in the north-west is evident. This gradient was removed by
subtracting a twisted plane from the map for the spectral
analysis. }
\label{i408}
\end{figure*}

Previous spectral index analyses towards the Cygnus Loop have been carried
out by Green~(\cite{gre90}) and Leahy \& Roger~(\cite{lea98}).  These
are based on 408~MHz and 1420~MHz interferometric measurements made
with the DRAO telescope (Landecker et al.~\cite{lan00}) and 2695~MHz
data from the Effelsberg 100-m telescope (Keen etal.~\cite{kee73}).
The latter map was one of the first observations ever taken with the
Effelsberg telescope.  At that time a one-channel receiver with a
linear feed was used. To account for the influence of polarized
emission, observations taken at different feed position angles were
averaged. During the last thirty years the Effelsberg 100-m telescope,
its receivers and the processing software acquired significant
improvements.  These technological advances are clearly reflected in
differences in faint emission areas between the Keen et al. map from
1973 and the recent 2675~MHz map described below.  The same advances
apply to the interferometer data as well. In particular, the data
reduction techniques at DRAO are continuously improved and much more
tools to process data, to remove the artifacts of strong sources are
now available.  Therefore we have reprocessed the DRAO 408~MHz and
1420~MHz data utilizing these new techniques.

\begin{figure*} 
\vspace{8cm}
\caption{Reprocessed 1420 MHz maps of the Cygnus Loop from the DRAO
archive. Short spacings are provided using the Uyan{\i}ker et
al.~(\cite{uya99}) Effelsberg maps, shown in Fig.~\ref{i1420}.  Left:
Contour plot representation. Note that in this panel the image is
convolved to $3\arcmin\times3\arcmin$ angular resolution to accentuate
the weak emission.  The first contour set on the left panel starts at
0.3 K T$_{\rm B}$ and runs in 0.25 K T$_{\rm B}$ steps.  The second
set starts at 1 K T$_{\rm B}$ and runs in steps of 0.5 K T$_{\rm
B}$. The third set is plotted between 4 and 12 K T$_{\rm B}$ with 2 K
T$_{\rm B}$ intervals.  Right: Grayscale image at full angular
resolution.  The HPBW is $1\arcmin\times2\arcmin$ (EW $\times$ NS).
}
\label{ic21}
\end{figure*}

The details of the 863~MHz and 1400~MHz Effelsberg maps are given by
Reich et al.~(\cite{rei03}) and Uyan{\i}ker et al.~(\cite{uya98},
\cite{uya99}), respectively.  Relevant observational parameters of
Effelsberg and DRAO data are summarized in
Tables~\ref{effdata}-~\ref{draodata}.  An outline of the new 2675~MHz
observations and reprocessing of DRAO archival data is given below.

\subsection{2675 MHz data}

We used the 2.7~GHz receiving system in the secondary focus of the
Effelsberg 100-m telescope, which was installed in 1997. The receiver
is a dual-channel polarization system recording left and right-hand
circularly polarized intensities by using highly stable cooled
HEMTs. The circularly polarized channels are connected to an
IF-polarimeter for simultaneous recording of correlated Stokes U and Q
data, and the system can be tuned to cope with the interference.  Our
observations were all made at 2675~MHz\footnote{ Note that Uyan{\i}ker
et al.~(\cite{uya02}) quote the upper band-limit of 2695~MHz instead
of the center frequency of 2675~MHz.}  with a bandwidth of 40~MHz. The
half power beam width (HPBW) of the 100-m telescope is $4\farcm 3$ at
that frequency. We used 3C286 as a main calibrator assuming 10.4~Jy,
percentage polarization of $9.9\%$ at a polarization angle of
33\degr. The conversion factor to obtain main beam brightness
temperature is T$_{\rm B}$/$S$ = 2.51~K/Jy (Reich et
al.~\cite{rei84}).

\begin{table}
\caption{Observational parameters of Effelsberg data} \label{effdata}
 \begin{tabular}{lcccc} \hline \hline
 Frequency  (MHz)    & 408  & 863  & 1400 & 2675  \\
 HPBW  (arcminute)   &  51  & 14.5 & 9.4  & 4.3   \\ \hline

 Calibrators         &      &      &      &       \\
 3C~286   (Jy)       &  -   & 18.5 & 14.4 & 10.4  \\
 3C~138   (Jy)       &  -   & 11.2 & 8.5  &  5.5  \\
 3C~48    (Jy)       & 38.9 & 22.4 & 15.9 &  -    \\

 References          & a, b & c    & d, e & f     \\ \hline
 \end{tabular}

 a) Haslam et al.~(\cite{has82}),
 b) Reich \& Reich~(\cite{rei88}),
 c) Reich et al.~(\cite{rei03}),
 d) Uyan{\i}ker et al.~(\cite{uya98}),
 e) Uyan{\i}ker et al.~(\cite{uya99}),
 f) Uyan{\i}ker et al.~(\cite{uya02})
 \end{table}

Observations of the Cygnus Loop were made on several nights between
March 1999 and August 2000. We mapped an area of $230\arcmin \times
270\arcmin$ centered on the Cygnus Loop in the equatorial coordinate
system by moving the telescope either along declination or right
ascension with a speed of 2\degr/min. The individual observations were
processed using the standard methods for continuum mapping
observations with the Effelsberg 100-m telescope. Linear baselines
were adopted by defining the data points at each end of a scan to
zero.  The edited individual observations were combined into maps for
the declination and the right ascension direction. In total about 6
coverages were obtained.  Finally these data were destriped and
combined using the {\sc plait} procedure (Emerson \& Gr\"ave
\cite{eme88}).  The final map is displayed in Fig.~\ref{i2675}.  The
projection of this map, as well as the subsequent maps, is in
rectangular RA/Dec coordinate system. Note that, due to the high
dynamic range of the images presented here, it is difficult to display
low and high emission features properly in the same plot. These
contrasts, however, clearly show up in color images or in electronic
display. We therefore display each image into two panels in contour
and grayscale representation.  This representation, thus, shows
various details of the structures, both fine and diffuse, and helps
comprehending the nature of the emission.  We measure an rms-noise of
$\sim$6~mK T$_{\rm B}$ from the total intensity image and $\sim$4~mK
T$_{\rm B}$ for Stokes U and Q.

We compared the new Effelsberg 2675~MHz map with the Keen et
al.~(\cite{kee73}) map. Sensitivity, enabling to discern the diffuse
emission around the remnant, is the most striking difference between
the two maps, though a general agreement between them does exist. Local
differences do not exceed 0.15~K T$_{\rm B}$, except the variable
sources which were reported by Keen et al.~(\cite{kee73}). The central
compact source CL4 ($\alpha, \delta\simeq\,$20$^{\rm h}$\, 50$^{\rm
m}$\, 48$^{\rm s}$, $31\degr\, 28^{\rm m}$), for instance, is about
0.7~K T$_{\rm B}$ stronger in the new map.

\subsection{DRAO data}

The DRAO data we have utilized have been extracted from the DRAO
archive, which were observed by Leahy et al.~(\cite{lea97}).  Due to the
recent improvements in the data reduction algorithms at DRAO, we have
reprocessed the 408~MHz and 1420~MHz synthesis telescope data. In this
way we have increased the frequency interval of Cygnus Loop
measurements to be used for a spectral analysis.

The new data reduction standards (Willis~\cite{wil99}) of the DRAO
synthesis telescope have been the topic of several papers explaining
the telescope (Landecker et al.~\cite{lan00}) and the Canadian
Galactic Plane Survey (CGPS, Taylor et al.~\cite{tay03}). These
methods now include procedures of editing, calibration, {\sc
clean}'ing and self-calibration as well as removing the grating
response of strong point sources inside and outside the field of view.
This is probably the most appreciable improvement.  The removal of
grating ring effects from strong sources makes a clear difference.  In
the past cleaned images of DRAO fields often contained residual
grating rings due to strong sources outside or inside the field, which
do not disappear when applying the standard {\sc clean}'ing process.
This happens because the clean components are subtracted by a Fast
Fourier Transformation procedure. Consequently aliasing artifacts
remain.  The strategy to remove the grating rings, in brief, is as
follows: we compute model visibilities from the best set of clean
components and then subtract the resulting model visibilities from the
set of visibilities which produced the best image.  Then we prepare an
image from the resulting visibilities and {\sc clean} this image.  The
resultant image is a {\sc clean}'ed map degraded to a minimum by
grating rings. That way an rms value of about 30~mK T$_{\rm B}$ is
measured in the mosaicked DRAO images. This is a particular
improvement in comparison to the previous method, where the removal of
grating rings was limited to the sources outside the main field.

\begin{center}
\begin{table}
\caption{Observational parameters for the DRAO data
} \label{draodata}
 \begin{tabular}{lcc} \hline \hline
 Frequency  (MHz)    & 408  &  1420            \\
 HPBW  (arcminute)   & 3.4$\times$6.9   &  1$\times$2  \\ \hline
 Calibrators         &      &                  \\
 3C~286   (Jy)       &  -   &  14.7            \\
 3C~295   (Jy)       & 54.0 &  22.1            \\
 3C~48    (Jy)       & 38.9 &  15.7            \\
 References          & a    &   a             \\ \hline
 \end{tabular}

  a) Landecker et al.~(\cite{lan00})
 \end{table}
 \end{center}

\begin{figure*}\centering 
\vspace{8cm}
\caption{
Sketch of continuum emission showing the major features towards
the Cygnus Loop. The outer diffuse plateau is shown by dotted
lines and the southern remnant is with dashed lines. Solid lines
mark the northern remnant.
}
\label{sketch}
\end{figure*}

Additionally we combined the DRAO synthesis telescope data at 1420~MHz
with recent Effelsberg data (Uyan{\i}ker et al.~\cite{uya99}).  For
this combination both DRAO and Effelsberg data sets are adjusted to
the temperature scale at 1420~MHz, which is tied to the 14.7 Jy flux
density of 3C~286.  This however is not a very significant correction
in terms of the resulting spectral indices. The effect of the
different temperature scales on the spectral index would introduce an
error 2\%, whereas the frequency difference of 20~MHz between the two
maps would cause an error of about 1\%. Nevertheless, to take this
difference into account the 1400~MHz Effelsberg map is adjusted to
1420~MHz scale assuming a spectral index value of $\alpha=-0.65$ for
3C~286 (Ott et al.~\cite{ott94}), before combining with the higher
resolution map.  The previous DRAO 1420~MHz map by Leahy et
al.~(\cite{lea97}) was corrected by Stockert 1.4 GHz (Reich et
al.~\cite{rei82}) data only.  Obviously, the Effelsberg--DRAO
combination has a better $u-v$ overlap in terms of baselines, and
therefore is superior to the older map.  Accordingly, the DRAO 408~MHz
map includes the Haslam et al.~(\cite{has82}) all-sky map at the same
frequency.  Note that 408~MHz Effelsberg data used for this analysis
take into account the temperature scale revision by Reich \&
Reich~(\cite{rei88}).  Thus we have obtained highest quality images to
date at these frequencies.  Images prepared in this way at 408~MHz and
1420~MHz are displayed in Fig.~\ref{i408} and Fig.~\ref{ic21}.

The difference image, prepared by subtracting the new DRAO images from
the previous images, is dominated by grating rings, which are
significant in low-level regions, but also influence strong emission
features.  In some places the intensities of these artifacts exceed
the intensities of diffuse regions. A more subtle problem is that the
grating rings from compact sources show up at different positions in
the 408~MHz and the 1420~MHz map.  Therefore, a spectral analysis
carried out with such images imposes severe limitations on its
accuracy.

In our final data set the Effelsberg observations at 863~MHz, 1400~MHz
and 2675~MHz have resolutions 14\farcm 5, 9\farcm 3 and 4\farcm 3,
while the DRAO data at 408~MHz and 1420~MHz are of $3\farcm 4\times
6\farcm 9$ (EW $\times$ NS) and $1\arcmin\times 2\arcmin$ (EW $\times$
NS), respectively. In all DRAO data used in this work short spacings
data from the Effelsberg telescope at the corresponding frequency are
incorporated, and hereafter by DRAO data we simply mean short spacings
corrected DRAO data.

\subsection{The radio appearance}

Radio emission from all the maps presented here commonly display the
two prominent shells, NGC~6960 and NGC~6992, the central filament,
southern remnant, a diffuse plateau predominantly encompassing the
northern remnant and a diffuse extended weak emission at the west of
the southern remnant.  All these major features are schematically
shown in Fig.~\ref{sketch}.

The plateau emission in the north encompasses the whole northern
remnant and positionally coincides with the reported Balmer-dominated
filaments (e.g., Hester et al.~\cite{hes94}). Therefore the boundary of
the plateau emission traces regions where the shock is progressing.

All radio maps reveal a diffuse  plateau enveloping the conventional bright
radio appearance of the Cygnus Loop. This plateau is the radio
counterpart to X-ray and infrared emission and is prominent towards
the north and north-eastern shell of the source.  This diffuse plateau
emission shows up in all of the recent Effelsberg maps at 863~MHz,
1400~MHz and 2675~MHz, but is just weakly represented in the original
interferometer maps. However, after combining the interferometer maps
with the single antenna data, the mottled boundary of the emission
plateau can be recognized.  Due to the limited coverage attained by
three individual interferometer fields the plateau emission,
especially towards east, stays very close to the edge of the
field. Similarly, this emission in the single-antenna data at 408~MHz
is overwhelmed by the existence of a strong large-scale gradient
extending from the north-west of the image down to the south-east
crossing over the whole Cygnus Loop complex.

\begin{figure*} 
\vspace{8cm}
\caption{ Spectrum of Cygnus Loop obtained from the maps given in this
paper (left) by integrating the emission from the regions enclosed by
the same polygon. The right panel shows the spectrum selected from all
available flux density values from the literature including those
obtained in this work.  The selection criteria are explained in the text.
The errors of the fits are at 3$\sigma$ level}.

\label{int}
\end{figure*}

\begin{table}
\caption{Integrated flux density values towards Cygnus Loop
} \label{int_table}
\begin{center}
 \begin{tabular}{rrcrc} \hline \hline

\multicolumn {1}{c}{ Frequency } &
  \multicolumn {3}{c}{ Flux density} &
   \multicolumn {1}{c}{ Ref} \\
  \multicolumn {1}{c}{ MHz } &
   \multicolumn {3}{c}{ Jy } &
    \multicolumn {1}{c}{  }  \\ \hline

  22 & 1378  &$\pm$& 400   & a                \\
 34.5& 1245  &$\pm$& 195   & b                \\
  38 &  956  &$\pm$& 150   & c                \\
  41 &  770  &$\pm$& 140   & d                \\
 158 &  350  &$\pm$&  70   & e                \\
 195 &  382  &$\pm$&  60   & d                \\
 408 &  230  &$\pm$&  50   & ~~c$^1$          \\
 408 &  260  &$\pm$&  50   & e                \\
 408 &  237  &$\pm$&  24   & ~~f$^2$          \\
 430 &  297  &$\pm$&  50   & d                \\
 863 &  184  &$\pm$&  18   & f                \\
 960 &  190  &$\pm$&  50   & ~~c$^3$          \\
1420 &  143  &$\pm$&  14   & f                \\
2675 &  115  &$\pm$&  12   & f                \\
2695 &  125  &$\pm$&  16   & ~~g$^4$          \\
2700 &   88  &$\pm$&   6   & h                \\
4940 &   73  &$\pm$&   7   & i                \\  \hline
\end{tabular}
\end{center}

 a) Roger et al.~(\cite{rog99});
 b) Sastry et al.~(\cite{sas81});
 c) Kenderdine~(\cite{ken63});
 d) Kundu \& Velusamy~(\cite{kun67});
 e) Mathewson et al.~(\cite{mat61});
 f) this work;
 g) Green~(\cite{gre90});
 h) Kundu~(\cite{kun69});
 i) Kundu \& Becker~(\cite{kun72});

$^1$ Flux density value revised by Roger et al.~(\cite{rog73})

$^2$ An arbitrary 10\% error in the integrated flux densities
to reflect the maximum possible error is assumed.

$^3$ Based on an earlier flux determination by Harris~(\cite{har62}).

$^4$ Obtained from the Keen et al.~(\cite{kee73}) map.
\end{table}

\section{Spectral analysis}
\subsection{Integrated flux density }

Although it is sometimes difficult to obtain accurate flux densities
for large SNRs, an accurate integrated spectrum is an important
characteristic of the object. Flux density integrations may depend on
the background adapted and on the selected emission region of the
source to be integrated.  Data at several frequencies treated in the
same way are needed to determine the overall spectral behavior. Such
an approach is adequate for the data considered in this paper.

Since there is a difference in the resolution at different wavelengths,
some faint point sources are not detected on low-frequency maps; this
is especially valid for the 863~MHz map.  For consistency, point sources in
the images are not excluded from the integrated flux density
calculations.

\begin{figure*}[htb] 
\vspace{8cm}
\caption{Selected rectangular regions used for the TT-plot analysis,
overlaid on top of the 2675~MHz total intensity map. The circles
denote regions including point sources, whose data were excluded from
the analysis.
}
\label{box}
\end{figure*}

Figure~\ref{int} shows the spectrum of Cygnus Loop obtained by
integrating the total-intensity images shown in the previous section.
The resultant integrated spectral index, fitted by taking the
observational errors given in Table~\ref{int_table} into account, is
$\alpha=-0.42\pm0.06$; where 0.06 reflects the 3$\sigma$ error. The
maximum deviation of the actual data points from the fit is about 3\%
in terms of the flux density, indicating that the adopted 3$\sigma$
level is a highly conservative estimate.

It is also useful to note the spectral index values calculated between
the pairs of integrated flux densities, as these values will provide a
reference to test the possible curvature across the source
(Sect.~\ref{ttplot}).  The integrated spectral index for the first
pair (between 1420 and 2675~MHz) is $\alpha=-0.34$. For the
frequencies 408 and 2675~MHz we obtain $\alpha=-0.38$, and for the 408
and 1420~MHz pair we get $\alpha=-0.40$.  These differences in the
spectral indices are likely due to the residual systematic scaling
errors not exceeding a few percent.

Table~\ref{int_table} is a compilation of the integrated flux values
for the Cygnus Loop, including the ones obtained in this work.  We
further inspect all of the published flux density values up to date in
order to extend the frequency coverage of the analysis.  Kovalenko et
al.~(\cite{kov94}), for instance, also give flux density values for
Cygnus Loop and suggest a break in the spectra. Although we have made
use of their compilation, we did not include data with inconsistent
baselevels, different scaling factors and those which have
insufficient resolutions almost comparable to the size of the
object. Since not all the flux integrations quoted in the literature
cover exactly the same areas, some scatter of the flux densities is
expected.  Nevertheless, a joint analysis of all available
flux-density values reveals an integrated spectral index of
$\alpha=-0.50$ with a larger scatter in the flux densities, compared
to the Effelsberg/DRAO values.  This is unavoidable when the
difficulties to integrate low-resolution measurements are taken into
account. However, the spectral index is quite close to the value
obtained from the present maps.  Most important of all, there is no
indication of any sort of spectral break in the spectrum. Such a break
was previously reported by DeNoyer~(\cite{den74}) with a break
frequency of about 1~GHz. This break likely reflects limited
sensitivity of some early measurements at higher frequencies resulting
in missing integrated flux density. This is another reason why some of
the earlier data are not included when calculating the present
spectrum.

\begin{figure*}\centering 
\vspace{8cm}
\caption{TT-plot analysis toward selected regions, shown in
Fig.~\ref{box}, between 1420~MHz DRAO data and 2675~MHz Effelsberg
data (TT1: pair of highest resolution maps). Labels at the lower right
of each plot correspond to those in Fig.~\ref{box}. The temperature
spectral index, $\beta$, obtained from this analysis is also
given. Note that $\beta = \alpha -2$.
\label{tt1}
}
\end{figure*}

\begin{figure*}\centering 
\vspace{8cm}
\caption{Same as Fig.~\ref{tt1}, but between 408~MHz DRAO data
and 2675~MHz Effelsberg data (TT2: widest frequency
separation)}
\label{tt2}
\end{figure*}

\begin{figure*}\centering 
\vspace{8cm}
\caption{Same as Fig.~\ref{tt1}, but between 1400~MHz
and 2675~MHz Effelsberg data (TT3: highest angular
resolution single antenna maps)}
\label{tt3}
\end{figure*}

\begin{figure*}\centering 
\vspace{8cm}
\caption{Same as Fig.~\ref{tt1}, but between 408~MHz DRAO data and
1420~MHz DRAO data (TT4). }
\label{tt4}
\end{figure*}

\subsection{TT-plots} \label{ttplot}

As demonstrated in the previous section, the integrated flux-density
values of the Cygnus Loop when measured in a wide frequency interval
at several frequencies reveal a fairly reliable spectral behavior for
the object. The obtained spectral index, $\alpha = -0.42$,
is rather typical for the majority of SNRs.

The method of differential spectral-index plots (TT-plot) provides an
alternative to flux integration, by plotting the temperature values of
images at two frequencies and fitting the resultant distribution with
a linear fit for dependent and independent variables separately --
where the average of the slopes corresponds to the temperature
spectral index, $\beta$ and the difference of the slopes gives the
nearly maximum possible error, $\Delta\beta$, of the spectral index.
This method is less dependent on remaining background emission,
particularly if the background emission is constant across the source.
For very large SNRs like the Cygnus Loop this assumption is
questionable, but holds for selected areas. Spectral results from
TT-plots should agree with the spectrum from integrated flux
densities, but are also useful to investigate local deviations from
the integrated spectrum. Note that the temperature spectral index,
$\beta$ obtained from a TT-plot, is related to the flux spectral
index, $\alpha$, as $\beta = \alpha -2$.

\begin{table*} \centering
\caption{Temperature spectral index values calculated for selected
regions as shown in Fig.~\ref{box}.
``Eff'' denotes Effelsberg measurements. All
frequency units are in MHz and all error values are rounded to the
nearest 0.01.}
\begin{tabular}{rcccc} \hline \hline
\multicolumn {1}{c}{ Region } &
 \multicolumn {1}{c}{$\beta$ (TT1) } &
  \multicolumn {1}{c}{$\beta$ (TT2) } &
   \multicolumn {1}{c}{$\beta$ (TT3) } &
    \multicolumn {1}{c}{$\beta$ (TT4) }  \\
\multicolumn {1}{c}{  } &
 \multicolumn {1}{c}{1420 DRAO -- 2675 Eff} &
  \multicolumn {1}{c}{408 DRAO -- 2675 Eff } &
   \multicolumn {1}{c}{1400 Eff -- 2675 Eff } &
    \multicolumn {1}{c}{408 DRAO -- 1420 DRAO }  \\ \hline

 1 &$-2.34\pm 0.06$ &$-2.38\pm 0.04$  &$-2.47\pm 0.08$ &$-2.40\pm 0.06$\\
 2 &$-2.27\pm 0.03$ &$-2.35\pm 0.01$  &$-2.43\pm 0.09$ &$-2.39\pm 0.02$\\
 3 &$-2.45\pm 0.14$ &$-2.39\pm 0.06$  &$-2.51\pm 0.07$ &$-2.36\pm 0.06$\\
 4 &$-2.43\pm 0.05$ &$-2.49\pm 0.03$  &$-2.29\pm 0.03$ &$-2.52\pm 0.03$\\
 5 &$-2.53\pm 0.13$ &$-2.44\pm 0.05$  &$-2.60\pm 0.06$ &$-2.38\pm 0.05$\\
 6 &$-2.46\pm 0.02$ &$-2.44\pm 0.01$  &$-2.52\pm 0.01$ &$-2.43\pm 0.02$\\
 7 &$-2.35\pm 0.03$ &$-2.38\pm 0.02$  &$-2.46\pm 0.01$ &$-2.39\pm 0.04$\\
 8 &$-2.39\pm 0.04$ &$-2.39\pm 0.01$  &$-2.56\pm 0.01$ &$-2.38\pm 0.02$\\
 9 &$-2.39\pm 0.09$ &$-2.37\pm 0.01$  &$-2.60\pm 0.05$ &$-2.34\pm 0.07$\\
\hline
\end{tabular}
\end{table*}

We have applied the TT-plot correlation method for a number of
selected regions of the Cygnus Loop.  The positions of the selected
sub-regions are shown in Fig~\ref{box}. We did not include the
Effelsberg 863~MHz map in this analysis because of its low angular
resolution of 14\farcm 5.  We used various pairs of maps at two
frequencies for TT-plots as follows:

\begin{itemize}
\item TT1:  1420~MHz DRAO and 2675~MHz Effelsberg maps, offering
 the highest angular resolution of $4\farcm 3$.

\item TT2: 408~MHz DRAO and 2675~MHz data, providing the widest
frequency separation. The angular resolution is $6\farcm 9$.

\item TT3:  1400~MHz and 2675~MHz Effelsberg maps, being
the highest resolution single-antenna data at $9\farcm 35$
angular resolution.

\item TT4: 408~MHz and 1420~MHz DRAO data, both being measured with an
interferometer and corrected for large-scale emission by incorporating
Effelsberg data as described above. The angular resolution is
$6\farcm 9$.
\end{itemize}

In order to avoid excessive oversampling, images are retabulated to a
grid of 2\arcmin. For each pair the higher resolution map is
convolved to the resolution of the lower resolution map.  By
interchanging the dependent and independent variables we have obtained
two $\beta$ values for each pair and the mean value of these fit
results is adopted as the temperature spectral index.  The difference
between the two $\beta$'s reflect the nearly maximum possible error
and taken as the error in the spectral index. Note that
Green~(\cite{gre90}) suggests another method to calculate the fit
errors for oversampled data by scaling with $\sqrt{\theta^2/(\Delta
x)^2}$, where $\theta$ is the half power beam with and $\Delta x$ is
the sampling of the data. Application of this correction for the data
at hand always yielded smaller error values for the spectral index
than the method used here. Therefore the more cautious
$\beta$-differences method for errors is kept.  Regions around point
sources, whose positions are examined in the highest resolution map at
1420~MHz from DRAO, are excluded from the TT-plot analysis, as their
contribution to the fitted slope will introduce a bias from the
spectrum of extragalactic sources.

TT1 (1420~MHz DRAO+Effelsberg -- 2675~MHz, Fig.~\ref{tt1}) provides
the most detailed spectral information across the remnant with a
better angular resolution than the other combinations. However, the
frequency interval is small. All of the resultant plots
give nonthermal spectral indices with a mean value
of $\alpha=-0.40$. This is  close to the spectral index calculated
from the integrated 1420~MHz and 2675~MHz values ($\alpha=-0.33$).
Both of these values are slightly above the overall spectral index
$\alpha=-0.50$.  The maximum spectral index difference is
$\Delta\alpha= 0.26$ between Box~\#2 and Box~\#5.

The largest frequency spacing has TT2 (Fig.~\ref{tt2})
(408~MHz--2675~MHz).  TT2 spectra are on average steeper when compared
to TT1.  The mean spectral index is $\alpha=-0.40$, close to
$\alpha=-0.38$ obtained from the integrated 408~MHz and 2675~MHz data.
There is less variation in $\alpha$ at different portions across the
source. The maximum spectral index difference is $\Delta\alpha= 0.14$
between Box~\#2 and Box~\#4.

TT3 (Fig.~\ref{tt3}) utilizes the two Effelsberg maps at 1400~MHz and
at 2675~MHz.  Here we obtain in general similar spectra as for
TT1. The mean spectral index is $\alpha=-0.50$. Differences between
TT3 and TT1 are largest for Box~\#8 and Box~\#9 ($\Delta\alpha =$ 0.17
and 0.21, respectively).  As for TT1 the maximum spectral index
difference is between Box~\#2 and Box~\#5 with $\Delta\alpha = 0.17$
in this case.

Analysis of the pair of interferometer maps at 408~MHz and 1420~MHz
with added short spacings provides a relatively wide frequency
coverage.  In this case (TT4, Fig.~\ref{tt4}) the mean spectral index
is $\alpha=-0.40$, exactly the same as that from the integrated flux
density values between these two pairs ($\alpha=-0.40$).  The maximum
spectral index difference $\Delta\alpha = 0.13$ is measured between
Box~\#2 and Box~\#4.

In general average values obtained from TT-plots agree very well with
the integrated values between the frequency pairs. This indicates that
integrated flux densities are not much affected by baselevel
uncertainties, for example by a warped background or ill-defined
outer boundaries of the sources.  However, as mentioned earlier, small
scaling errors of a few percent might cause small systematic
differences between the different frequency pairs.  Therefore, this
scatter should not be taken at face value for the spectral index
changes.

\begin{figure*}
\vspace{8cm}
\caption{Temperature spectral index maps between different frequency
pairs. at 14$\farcm$5 resolution. Contours start at -2.8 and run in
steps of 0.1. White contours start at -2.4 and run in steps of 0.1.
}
\label{beta}
\end{figure*}

It is interesting to compare the present results between 408~MHz and
2675~MHz (TT2) with those of Green~(\cite{gre90}) and Leahy \&
Roger~(\cite{lea98}) for the same frequency interval.  On average
Green's spectra agree exactly with the present ones ($\alpha=-0.41$),
but the scatter of his values is larger ($\Delta\alpha = 0.36$
compared to $\Delta\alpha = 0.18$ in TT2).  The analysis of Leahy \&
Roger gives an even larger scatter for the same frequency interval
with $\Delta\alpha = 0.48$.  We have tried to analyze the reason for
this discrepancy by checking individual areas for systematics.  For
the strong north-eastern filament NGC~6992 (Box~\#1 and Box~\#2)
Green's result is rather similar to the present one, while Leahy \&
Roger obtained spectra which are about $\Delta\alpha \sim 0.2$
flatter. For the southern remnant (Box~\#8 and Box~\#9) the agreement
with Leahy \& Roger is within $\Delta\alpha \sim 0.1$, while Green's
spectra are flatter towards the east and steeper to the
west. Obviously the differences vary from region to region, and we
conclude that the reduced scatter in the spectral indices derived in
the present study are based on an improved data quality of the 408~MHz
and the 2675~MHz maps. The steepest spectrum is observed for Box~\#5
($\alpha=-0.53$), where the emission level is relatively low; the
above mentioned studies also obtained similar values in this region.
As noted above, the 408~MHz map shows a background emission gradient
across the map, which was subtracted by a twisted plane. However, an
influence of residual steep-spectrum Galactic background emission
might influence in particular spectral indices for low emission
regions like in Box~\#5. Without Box~\#5 the spectral index scatter
reduces to $\Delta\alpha \sim 0.11$.

The results of Leahy \& Roger for spectral index variations between
408~MHz/1420~MHz and 1420~MHz/2695~MHz are significantly larger than
the present ones. For TT-plots between 408~MHz/1420~MHz they report
spectra between $\alpha=+0.11$ and $\alpha=-0.80$ and for
1420~MHz/2695~MHz between $\alpha=0.0$ and $\alpha=-0.70$. This large
variation is not confirmed by the present analysis and we believe the
discrepancy is due to the limited quality of the earlier maps used.

The main sources of errors in this analysis are due to the calibration
of the maps and background determination. An error of about 10\% in
the individual maps on the other hand causes an uncertainty of
$\pm$0.06 within $3\sigma$ confidence interval. All of the errors from
the TT-plot results remain within the $3\sigma$ limit (exception to
this are boxes \#3, \#5, and \#9 from TT1 and boxes \#1, \#2, and \#3
from TT3).  If the individual spectral indices are compared with the
overall index $-0.42\pm0.06$, we find that values from Boxes \#6 to
\#9 systematically deviate from the overall index more than the errors
of the fits.  The values from the other boxes are, in spite of some
scatter, less than or comparable with the errors. The exception here
again is the Box\#4---which is most likely affected by the spectra of
the large-scale diffuse emission.  Excluding Box\#4, Boxes \#6 to \#9
all show steeper spectra deviating from the overall spectra about
$\Delta\alpha=0.1$.

\subsection{Spectral index maps between pair of frequencies}

The spectral index variations at different frequency pairs can be
followed best in spectral index maps. Such maps are given in
Fig.~\ref{beta}. In preparing these images all maps are convolved to a
resolution of $14\farcm 5$ and based on the result of a TT-plot
analysis of the entire source zerolevels of the individual images are
adjusted.  According to this scheme 600, 0, 23, and 10 mK offsets are
added to the images and to assure high signal to noise ratio, pixel
values less than 6~K, 1~K, 0.3~K, and 65~mK are blanked, at 408, 863,
1420, and 2675~MHz, respectively.

Obviously, low intensity regions are more sensitive to any zerolevel
and scaling uncertainties.  This effect is clearly visible towards the
edges of the object and in the lowest intensity region between
NGC~6992 and the central filament, where spectra are as steep as
$\alpha=-0.7$.  The spectral index map for 863/1420~MHz has the
smallest frequency interval and therefore has an irregular appearance
in comparison to the other pairs; with increasing frequency separation
the characteristics of the spectral index distribution becomes
conspicuous. Leaving this map aside, all spectral index maps shown in
Fig.~\ref{beta} show the three flat spectrum regions, towards the
filaments, and one steep spectrum region towards the center. The
spectra obtained from these maps slightly differ from those of TT-plot
analysis. However one thing here must be kept in mind: the TT-plot
analysis is sensitive to fluctuations rather than large-scale
emission. Nevertheless, the tendency in the spectra in both analyses
is similar. The Boxes~\#4, \#5, and \#6 corresponding to the steep
spectra region in the spectral index maps, for instance, also have
steep spectra in TT-plot analysis.

Three regions with flat spectra (towards north-east, north-west and
south) are prominent in these maps, whereas in the central portions of
the object the spectra become steeper.  There seems to be at least two
major spectral regions with $\alpha\simeq-0.4$ towards the northern
part and $\alpha\simeq-0.6$ towards the southern part of the Cygnus
Loop, both confined by the flat spectrum filaments.

\subsection{Spectral index map between four frequencies}

In order to obtain further and overall spectral information on the
Cygnus Loop we used all maps available.  These are at 408~MHz,
863~MHz, 1420~MHz and 2675~MHz at an angular resolution of $14\farcm
5$. By means of a linear fit to the four temperature values from each
map, for every pixel, we have determined the overall average spectral
index. Intensities lower than the limits given in the previous section
are flagged invalid and these regions are blanked. Intercepts of the
fitted lines are also stored in an image to check inconsistent fits;
an inspection of this image confirmed that background levels of the
images were properly set and the gradient in the 408~MHz map
adequately removed. The resultant map of spectral indices is displayed
in Fig.~\ref{spec}.  Further, we have calculated the residual maps of
differences from the fitted spectral indices for each map and found no
systematic deviations indicating scaling errors or spectral bends.
The map shows spectral index variations between $\beta=-2.50$ and
$-2.30$. In the weak, extended emission regions of the Cygnus Loop the
spectra are, in general, steeper; which reflects the background
emission on the lower frequency images. The span of the temperature
spectral index obtained from this analysis is about the same range as
those obtained from the TT-plots and simply indicates small effects on
spectral indices by varying Galactic background emission.

\begin{figure*}
\vspace{8cm}
\caption{Temperature spectral index map calculated from maps at four
frequencies between 408~MHz and 2675~MHz at $14\farcm 5$ angular
resolution.  The grayscale extends from $-$2.5 (light) to $-$2.3
(dark).  The overlaid contours start from $-2.5$ and run in
steps of 0.05, where the  white contours start from $\beta=-2.40$.
}
\label{spec}
\end{figure*}

There are, however, three distinct regions seen in this image.
Towards NGC~6992, the region to the north-eastern edge of the northern
remnant and the southern part of the southern remnant all show higher
spectral indices with values around $\alpha\simeq-0.4$.  Interiors of the
remnant has slightly steeper spectra, where the intensities are
lower.  These differences agree with those from TT2, which have the
largest spacing in frequency. Both methods give about the same range
of differences in the spectral indices, which do not exceed
$\Delta\alpha=0.2$.  The spectra from TT1 and TT3, with smaller
frequency separation, are limited in accuracy by residual systematic
effects.

>From our spectral index results for the Cygnus Loop we conclude that
spectral index variations are small. We do not find evidence of a
spectral break in the frequency range up to 2675~MHz.  It is not
entirely clear to what extent the remaining spectral index variation
depends on residual effects from calibration uncertainties, unresolved
background source contributions and limitations when matching maps
from different telescopes with different angular resolutions. To
overcome these limitations special care is needed when planning
dedicated observations for the purpose of spectral index work.

\section{Remarks on spectral variations}

In general, radio spectra of SNRs can be fitted rather well by a
power-law.  Most SNRs are likely in the adiabatic phase of
evolution. Strong shocks with a compression ratio $\sim4$ accelerate
electrons with a synchrotron emission spectrum of $\alpha=-0.5$. The
overall Cygnus Loop spectrum is rather close to this value. Although
we clearly rule out large spectral index changes, small spectral
variations are indicated.  Some effects are known which may cause
deviations in the spectra, which in principle can be effects occurring
internal to the source or are external effects due to the intervening
interstellar medium. Spectral variations are therefore of high
interest and were already discussed in detail by previous workers
(e.g. Caswell et al.~\cite{cas71}; Lang~\cite{lan74};
Kassim~\cite{kas89}; Anderson \& Rudnick~\cite{and93}; and Leahy \&
Roger~\cite{lea98}). Here we briefly summarize these mechanisms
and discuss their relevance to the Cygnus Loop.

Due to free-free absorption by foreground ionized material a turnover
at frequencies between 10~MHz and 100~MHz might result.  This seems
irrelevant for the Cygnus Loop as indicated by the flux density at
22~MHz (Roger et al.~\cite{rog99}), matching the high-frequency power
law.  Excessive thermal material should also significantly enhance
rotation measure (RM) values; such RMs, however, are not observed
(Uyan{\i}ker et al.~\cite{uya02}, the mean value of the rotation
measure is about $\rm -20\, rad\, m^{-2}$).  At a distance of 440~pc
(Blair et al.~\cite{bla99}), the linear size of the Cygnus Loop
complex would be about 25~pc.  The inferred thermal electron
densities, with $B_{\|}=3~\mu$G, are then of the order of 0.33
cm$^{-3}$. The corresponding emission measure is 2.7 pc cm$^{-6}$,
giving an excess emission of $\sim1$~mK at 2675~MHz; but this excess
emission is too weak to have any influence on the spectral shape.
Other absorption effects (Tsytovich-Razin effect and synchrotron
self-absorption) were shown by Leahy \& Roger~(\cite{lea98}) to be
irrelevant for the Cygnus Loop as well. Leahy \& Roger also considered
a spectral steepening by synchrotron energy losses and found
unrealistic time scales exceeding typical SNR lifetimes.  Otherwise
exceptional high magnetic fields need to be assumed and there is no
indication for that.  Note, however, that the strength of the shock is
expressed as the ratio of the densities before and after the shock
$r=\rho_1/\rho_0$. For nonrelativistic shocks $r=4$ and $\alpha=-0.5$.
Relativistic shocks produce $r=7$ (with $\alpha=-0.25$). Radio spectra
flatter than $\alpha=-0.5$ may, therefore, also be caused by slightly
relativistic shocks.

The spectrum of the Galactic background radio spectrum has a turnover
between 100 and 600~MHz (Webster~\cite{web74}) and can be shifted to
higher frequencies by compression of magnetic fields as discussed by
van der Laan~(\cite{van62}).  There are strong evidences that the
northern remnant and the most prominent filament at the upper part of
this remnant are aligned with the large-scale local magnetic field in
this region (Uyan{\i}ker et al.~\cite{uya02}).  The two brightest
partial shells, NGC~6960 and NGC~6992, may be formed due to magnetic
compression. It is interesting that slightly flatter spectra are
observed in these regions.  The effect of spectral variation due to
compression of the Galactic magnetic field was previously observed for
the SNR S147 (F\"urst \& Reich~\cite{fur86}). The filamentary
structures of S147 have significantly flatter spectra in the
GHz-frequency range than that of diffuse structures. This spectral
variation is of the order of $\Delta\alpha=0.4$.  S147 clearly shows a
spectral break in its spectrum above 1~GHz, and the influence by
differences in the compressed magnetic field are quite clear.  The
Cygnus Loop seems to be in an earlier phase of evolution if compared
to S147 where shock acceleration is still dominant. It would be quite
interesting to have high-quality maps at 5~GHz or even higher
frequencies to be included into the spectral investigation.

We observe complex variations of the spectral indices across the whole
Cygnus Loop.  The upper limit of these variations is
$\Delta\alpha=0.2$.  Considering the low errors obtained in the
TT-plot analysis and in the spectral index maps, these variations must
be real. The steepest spectral indices towards very low intensity
regions at the edges of the object, on the other hand, are not.

On the whole, all of the effects listed above may play some role in
shaping the spectra of the remnants, while some of them are
negligible. In general the results of the spectral analysis indicate
that the spectra of the Cygnus Loop  are quite typical for
shell-like remnants and are not extraordinary.

The spectral index variations across the Cygnus Loop and the quite
straight integrated spectrum are not in contradiction as we can show
by a simple model.  Based on a rough estimate we assume a fraction of
45\% of the Cygnus Loop's flux density at 2675~MHz to have
$\alpha=-0.3$, 25\% has $\alpha=-0.5$ and 30\% has $\alpha=-0.4$.
Extrapolating these components between 22~MHz and 10~GHz results in an
integrated spectral index varying from $\alpha=-0.396$ between 22~MHz
and 2675~MHz to $\alpha=-0.376$ between 2675~MHz and 10~GHz. Even for
a very wide frequency range significant spectral index variations
within the SNR do not reflect in the integrated spectral index when
realistic flux density errors are taken into account. This should be
considered in the discussion for distant barely resolved SNRs, where
only an integrated spectrum can be derived.

\section{Conclusions}

We presented the results of a spectral index analysis of the Cygnus
Loop based on a number of high-quality maps.  While the integrated
spectrum is rather straight with a spectral index of $\alpha = -0.42$,
we reveal spectral index variations, with an upper limit of
$\Delta\alpha=0.2$, which most likely reflect some weak influence of
the compressed Galactic magnetic field.  Previously reported curved
spectra varying across the source are not confirmed.

We found at least three spectral components across the Cygnus Loop,
corresponding to at least five different regions with flat and steep
spectra. The difference between the spectra of these regions reaches
up to $\Delta\alpha=0.2$. The north-western part of the object has
$\alpha\simeq-0.4$, except for the filaments where spectrum is even
flatter. For the southern part the spectrum is steeper, with
$\alpha\simeq-0.6$, again the filaments in this regions are
exceptional. With these spectral differences the southern part should
dominate the Cygnus Loop at low frequencies, while at high frequencies
the flat spectrum filaments are strongest and the steep-spectrum
central region weakens.  The flat spectrum filaments, placed on top of
a steeper spectrum object and implying high magnetic fields, are
likely similar to those observed towards S~147.

\begin{acknowledgements}
We like to thank Dennis Leahy and Rob Roger for providing us with
their archival DRAO Cygnus Loop observations for reprocessing and
Patricia Reich for critical reading of the manuscript.  We also
thank the anonymous referee for constructive comments. The Dominion
Radio Astrophysical Observatory is operated as a National Facility by
the National Research Council of Canada.
\end{acknowledgements}

\end{document}